\documentclass[aps,prl,superscriptaddress,amsmath,amssymb,preprintnumbers,twocolumn,showpacs]{revtex4}
\usepackage{graphicx,color}
\usepackage{amssymb}
\usepackage{amsmath}
\usepackage{bm}

\newcommand{\bra}[1]{\langle #1 |}
\newcommand{\ket}[1]{| #1 \rangle}

\definecolor{dgreen}{rgb}{0,0.5,0}

\definecolor{delete}{cmyk}{0.5,0,0,0}
\definecolor{deletey}{cmyk}{0,0.5,0,0}

\newcommand{\NTT}{NTT Basic Research Laboratories, NTT Corporation, 3-1
Morinosato-Wakamiya, Atsugi, Kanagawa, 243-0198, Japan.}
\newcommand{\waseda}{Department of Physics, Waseda University, Tokyo 169-8555, Japan}

\begin{document}


\title{
Robust entanglement-based magnetic field sensor
beyond the standard quantum limit
}
%


\author{\ \ \ \ \ \ \ \ \ \  \ \ \ \ \  \ \ \ \ \ \ \ \ \ \  \ \ \ \ \ \
\ \ \ \ \ \ \ \ \  Tohru Tanaka\footnote{These authors equally contributed to this paper.}}
\affiliation{\NTT}
\affiliation{\waseda}
\author{Paul Knott$^*$}
\affiliation{\NTT}
\author{Yuichiro Matsuzaki$^*$}
\affiliation{\NTT}
\author{\newline Shane Dooley}\affiliation{\NTT}
\author{Hiroshi Yamaguchi} 	\affiliation{\NTT}
\author{William J. Munro}				\affiliation{\NTT}
\author{Shiro Saito} 	\affiliation{\NTT}


\begin{abstract}
Recently, there have been significant developments in entanglement-based quantum metrology.
However,  entanglement is fragile against experimental imperfections,
and \textcolor{black}{quantum sensing to beat the standard quantum limit
 in scaling has not yet been achieved in realistic systems.}
Here, we show
that it is possible to overcome such restrictions
so that one can sense a magnetic field with an accuracy
beyond the standard quantum limit even under the effect of decoherence,
by using a realistic entangled state that can be easily created even with current technology.
Our scheme could pave the way for the realizations of practical entanglement-based magnetic field sensors.
\end{abstract}

\pacs{03.67.-a, 
03.75.Dg 
}

\maketitle

Precise measurement of a weak magnetic field is one of the important goals
for sensing technologies~\cite{metro_bio,review_q_metro_1,review_q_metro_2,wineland1992,wineland1994}.
Such measurement has potential applications
in the field of materials science, biology~\cite{metro_bio}, and foundations of physics~\cite{review_q_metro_1,review_q_metro_2}.
For the estimation of the magnetic field,
one usually prepares $N$ degenerate electron spins as a probe system~\cite{wineland1992,wineland1994}.
Since magnetic fields induce a finite energy shift of the electron spins,
the electron spins under the effect of the magnetic fields acquire a relative phase
when the state contains a superposition.
Therefore it is possible to estimate the magnetic field
exposed with the electron spin by measuring the relative phase.

In
quantum metrology~\cite{review_q_metro_1,review_q_metro_2},
when the probe composed of $N$ spins has no quantum correlations,
the uncertainty in the estimation scales as $O(N^{-\frac{1}{2}})$, the standard quantum limit (SQL)~\cite{review_q_metro_1,review_q_metro_2}.
On the other hand, by preparing the probe in an entangled states,
the estimation uncertainty can be in principle reduced to $O(N^{-1})$, the Heisenberg limit~\cite{review_q_metro_1,review_q_metro_2}.
Hence, there have been many efforts in both theory and experiment
regarding entanglement-based phase measurement~\cite{review_q_metro_1,review_q_metro_2,review_q_metro_3,review_q_metro_4,review_q_metro_5}.
Especially,
there are many theoretical studies about the effect of decoherence~\cite{mark_noise_1,mark_noise_2,mark_noise_3,mark_noise_4,mark_noise_5,non_mark_noise_1,non_mark_noise_2}.
It is known that, under the effect of Markovian dephasing,
the estimation uncertainty scales the same as the SQL~\cite{mark_noise_1,mark_noise_2,mark_noise_3,mark_noise_4},
which means that the entanglement does not provide any advantages over the classical strategy in scaling.
However,
it has been recently shown that the phase measurement with the GHZ state can actually beat the SQL
under the effect of some Markovian noise~\cite{mark_noise_5} or non-Markovian dephasing~\cite{non_mark_noise_1,non_mark_noise_2}.
Especially, the estimation uncertainty scales as $O(N^{-\frac{3}{4}})$ under the effect of non-Markovian dephasing~\cite{non_mark_noise_1,non_mark_noise_2}.
Since the relevant noise in solid-state systems
is usually non-Markovian dephasing~\cite{noise_solid_1,noise_solid_2,noise_solid_3,noise_solid_4,noise_solid_5,noise_solid_6,noise_solid_7,noise_solid_8,noise_solid_9},
this result opens a way to realize practical entanglement-based metrology.

It is difficult to generate a large $N$-qubit GHZ state because this usually requires $N$ operations.
Moreover, individual access to each qubit is typically needed for the creation of the GHZ state.
Although there are reports of small size GHZ states generated with current technology~\cite{cre_ghz_1,cre_ghz_2,cre_ghz_3},
an experimental demonstration to create a GHZ state in a scalable way has
not yet been done.
A large entangled state is necessary to construct a quantum sensor far
beyond the accuracy of classical sensors, and
so it is crucial for realizing the entanglement-based sensor
to pursue the possibility of using other types of entanglement that can be created in more efficient ways.

In this paper, we show a way to construct a robust entanglement-based quantum field sensor where
a large entangled state can be created and read out even under the effect of experimental imperfections.
Specifically, we investigate spin cat states and spin-squeezed states, which can be created with current technology.
There have recently been many theoretical and experimental studies
for the generation of spin cat states~\cite{sc_shane2014,sc_shane2014_2} and spin squeezed states~\cite{review_q_metro_4,squeezed_review,squeezed_experi_1}.
As explained below,
we can generate these states by ``global operations" such as the application of a microwave pulse to the ensemble and a collective interaction.
This means that the necessary number of operations is constant to generate entangled states of arbitrary size.
Moreover, we show that our quantum strategy beats the SQL in scaling even under the effect of realistic decoherence.

\begin{figure}[tb]
\centering
\includegraphics[width=72mm, height=36mm]{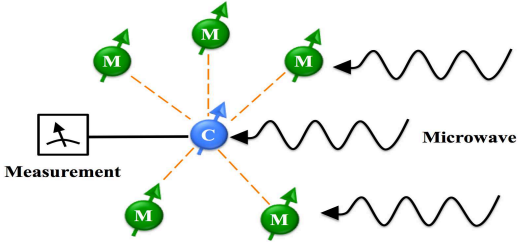}
\caption{(Color Online). Schematic diagram of our system. Long-lived
 memory qubits are coupled with a central control qubit. These
 qubits are manipulated by
 microwave pulses.}\label{schematic}
\end{figure}

{\it Entanglement as a resource--}
{Let us define a spin coherent state,  a spin-squeezed state, and a spin cat state.
The spin coherent state is defined as $\ket{z , N} = \frac{1}{({1 + |z|^2})^{N/2}} (\ket{g} + z \ket e )^{\otimes N}$,
where $\ket g (\ket e )$ is the eigenstate of $\sigma_z$ with an eigenvalue $-1 (+1)$ and $z$ is a complex number~\cite{squeezed_spin:1993}.
By preparing a spin ensemble in a spin coherent state
and letting it evolve by a one-axis (two-axis) twisting Hamiltonian~\cite{squeezed_spin:1993},
we obtain a one-axis (two-axis) twisted spin squeezed state as
$| \mbox{OAT} \rangle = e^{-i \chi J_z^2} \ket{z , N}, \quad |z| = 1,$ ($| \mbox{TAT} \rangle = e^{ \chi (J_+^2 - J_-^2)} \ket{0 , N}$)
where $\chi$ denotes a real number, $J_z= \frac{1}{2}\sum_{i=1}^N \sigma_{z,i}$ denotes a collective angular momentum operator for the spin ensemble,
and $ J_{\pm}=\sum_{i=1}^{N}\hat{\sigma }_{\pm }$ is a collective ladder operator.
A spin cat state is defined as
$\ket {\mbox{SC}} =\frac{1}{\sqrt{2}} (|0,N\rangle + |z,N\rangle) $ for $N\gg 1$.
Quantum sensing with spin cat states or spin squeezed states can beat the SQL without decoherence~\cite{footnote_1,qft_spin_cat,shanecat2013}.

{{\it System--}
Suppose that a short-lived and controllable qubit
is collectively coupled to $N$ long-lived
qubits as described in the Fig.~\ref{schematic}.
We call the former a control qubit and the latter memory qubits.
The memory qubits are used as a probe for the magnetic field while
the control qubit is used to generate entanglement between
the memory qubits and to read out the phase information acquired in the memory qubits.
The Hamiltonian is described as
\begin{equation}\label{hami}
\begin{split}
H&= H_{\text{c}}+H_{\text{m}}+ H^{(1)}_I + H^{(2)}_I\\
 H_{\text{c}}&=\frac{\omega
 _{\text{c}}}{2}{\sigma}^{(\text{c})}_z+\frac{\lambda _c}{2}\hat{\sigma
 }^{(\text{c})}_x \cos (w t +\phi ), \\
H_{\text{m}}&= \omega_{\text{m}} J_z^{(\text{m})} +\lambda
 _{\text{m}}\hat{J}_x \cos (w t +\phi ) \\
H^{(1)}_I&=  g_1{\sigma}^{(\text{c})}_z J_z^{(\text{m})}, \quad H^{(2)}_I= g_2({\sigma}^{(\text{c})}_+ J_-^{(\text{m})} + {\sigma}^{(\text{c})}_- J_+^{(\text{m})})  ,\nonumber
\end{split}
\end{equation}
where ${\sigma}^{(\text{c})}_x$ and ${\sigma}^{(\text{c})}_z$ denote Pauli  operators acting on the control qubit,
$J_x^{(\text{m})}$ and $J_z^{(\text{m})}$  denote collective angular momentum operators acting on the memory qubits,
$ {\sigma}^{(\text{c})}_{\pm}$ ($ J_{\pm}^{(\text{m})}$) is a ladder operator acting on the control (memory) qubit,
$\omega _{\text{c}}$ ($\omega _{\text{m}}$) denotes the energy of the control (memory) qubit,
$w$ denotes the microwave frequency, $\phi $ denotes the microwave phase,
$g_1$ $(g_2)$ is a coupling constant,
and $\lambda _c$ ($\lambda _m$) denotes the Rabi frequency of the control (memory) qubit.
We assume that with this system we can
(i) implement a projective measurement on the control qubit,
(ii) tune the resonant frequency of the control qubit,
(iii) switch the interaction Hamiltonian from $H_I^{(1)}$ to $H_I^{(2)}$ (and vice versa),
(iv) prepare the ground state of this system,
and (v) change the Rabi frequency and microwave phase in an arbitrary timing.

One of the experimental realizations of our setup is a hybrid system of a superconducting flux qubit and negatively charged Nitrogen-vacancy (NV) centers~\cite{marcos2010coupling, zhu2011coherent,saito2013towards}.
The superconducting flux qubit has excellent controllability for single qubit rotation, frequency control, and projective measurement~\cite{noise_solid_7,ClarkeWilhelm01a}.
On the other hand, the NV centers~\cite{footnote_M} typically have a long coherence time of hundreds of microseconds~\cite{maurer2012room}.
Since the collective coupling between the flux qubit and NV centres has been
experimentally demonstrated~\cite{zhu2011coherent,saito2013towards},
this is one of the promising systems to realize our theoretical proposal.
}

{{\it Magnetic field sensing--}}
We introduce our setup for estimating a magnetic field.
To include realistic imperfections,
we consider the effect of independent non-Markovian dephasing while the memory qubits are exposed to the magnetic field.
First, the memory qubits are prepared in an entangled state such as a spin cat state or spin-squeezed state $\ket \psi$.
Second, the target magnetic field $\omega$ is embedded as $\rho _t = e^{- i \omega t J_{\vec n}} {\mathcal E}^{\otimes N} (\ket \psi \bra \psi) e^{ i \omega t J_{\vec n}} ,$
where $\vec n$ is a three dimensional real vector with unit length
and ${\mathcal E}^{\otimes N}$ denotes an independent non-Markovian dephasing.
The action of ${\mathcal E}$ is defined as
\begin{equation}\label{noise_map}
\begin{pmatrix} a & b \\ c & d \end{pmatrix} \quad \stackrel{{\mathcal E}}{\to} \quad \begin{pmatrix} a & e^{-(\Gamma_tt)^2}b \\ e^{-(\Gamma _tt)^2} c & d \end{pmatrix},
\end{equation}
in a basis diagonalizing $\vec n \cdot \vec \sigma$ where $\Gamma _t$ denotes a time dependent decoherence rate.
This time dependent decoherence rate is known to be scaled as $\Gamma
_t=O(t^0)$ for a small $t$~\cite{noise_solid_1,noise_solid_2,noise_solid_3,noise_solid_4,noise_solid_5,noise_solid_6,noise_solid_7,noise_solid_8,noise_solid_9}.
Hence, we define $\gamma=\lim_{t\rightarrow 0}\Gamma _t$.

\textcolor{black}{
To calculate the precision with which our scheme can measure a magnetic field,
we use the quantum Fisher information ${F}(\rho_t)$, which does not depend on the magnetic field $\omega$ in the above setup.
Here, if we read out the field from an expectation value of an observable $A$, ${\mbox{E} }_{\rho_t} (A)$~\cite{footnote_2},
the inequality ${F} (\rho_t)\cdot  {\mbox{Var}}_{\rho_t} (A) \geq \bigg| \frac{\partial }{\partial \omega} {\mbox{E} }_{\rho_t} (A) \bigg|^2$
holds for any $\omega$~\cite{PhysRevA.70.022327,qft_ref_2},
where ${\mbox{Var}}_{\rho_t} (A)$ is the variance of $A$.
From this we can find the estimation uncertainty of $\omega$ to be $\delta \omega = \sqrt{\frac{1}{\mu}}
\frac{\sqrt{\mbox{Var}_{\rho_t} (A)}} {|\frac{\partial}{\partial \omega}
{\mbox{E}}_{\rho_t} (A) |}$ where $\mu$ is the number of measurement data points.
We will approximate the number of measurement data points as $\mu
\simeq T/t$, where $T$ is a total measurement time.
This assumption is valid when the coherence time of the
memory qubits is much longer than any other times for operations.}

{{\it Entanglement sensor with spin cat states--}}
We show a new way to prepare the memory qubits in a spin cat state.
By selecting $H_I^{(1)}$ as the interaction Hamiltonian,
the state of the memory qubits (control qubit) can change the resonant frequency
of the control qubit (memory qubits) from $\omega _c$ ($\omega
_{\text{m}}$)  to $\omega_{\text{c}}+2g_1J_z$ ($\omega_{\text{m}}+g_1{\sigma }_z$).
We can use these properties to make the spin cat state.
First, we prepare a ground state for the memory qubit and a superposition of the controller qubit $\frac{1}{\sqrt{2}}(|g\rangle _{\text{c}}+|e\rangle_{\text{c}})|0,N\rangle _{\text{m}}$.
This superposition can be made by applying a $\frac{\pi }{2}$ pulse with a frequency of $\omega_{\text{c}}-g_1N$ on a ground state of the control qubit.
Next, we perform a selective pulse with a frequency of $\omega _{\text{m}}-g_1$ to rotate the memory qubits,
so that we obtain $\frac{1}{\sqrt{2}}|g\rangle _{\text{c}}|z,N\rangle_{\text{m}}+\frac{1}{\sqrt{2}}|e\rangle _{\text{c}}|0,N\rangle_{\text{m}}$.
Finally, we can perform a selective $\pi $ pulse with a frequency of $\omega_{\text{c}}-g_1N$ on the controller qubit
and obtain a spin cat state $|g\rangle _{\text{c}}\frac{1}{\sqrt{2}}(|0,N\rangle_{\text{m}}+|z,N\rangle _{\text{m}})$.


Let us now describe how to read out the phase induced by a
target magnetic field from the spin cat state.
We expose the sensor to the magnetic field for a time $t$, and
the spin cat state acquires
a relative phase $\omega t$ such that $|g\rangle _{\text{c}}\frac{1}{\sqrt{2}}(|0,N\rangle_{\text{m}}+|ze^{-i\omega
t},N\rangle _{\text{m}})$ due to the interaction with the magnetic field. 
To readout this phase, we apply two selective pulses, and
perform a projective measurement on the control qubit.
The first selective $\pi $ pulse (with a frequency of
$\omega_{\text{c}}-g_1N$) is applied on the control qubit,
giving $\frac{1}{\sqrt{2}}|e\rangle _{\text{c}}|0,N\rangle_{\text{m}}+\frac{1}{\sqrt{2}}|g\rangle _{\text{c}}|ze^{-i\omega t},N\rangle _{\text{m}}$.
The second selective pulse
(with the frequency of $\omega _{\text{m}}+g_1$) is applied on the memory qubits,
giving $\frac{1}{\sqrt{2}}|e\rangle _{\text{c}}|z,N\rangle_{\text{m}}+\frac{1}{\sqrt{2}}|g\rangle _{\text{c}}|ze^{-i\omega t},N\rangle _{\text{m}}$.
If we perform a projective measurement about ${\sigma }_y$ on the control qubit,
the probability to obtain a measurement result of ${\sigma }_y=+1$ is
given by
$P_{+}=\frac{1}{2}+\frac{1}{2}\text{Im}[\langle z|ze^{-i\omega t}\rangle]\simeq \frac{1}{2}-\frac{1}{2}\frac{|z|^2}{1+|z|^2}N\omega t$
for $N\omega t \ll 1$. Thus we can
estimate the phase from the measurement.
Note that this measurement can be approximated as a projective measurement onto $\ket{\Phi_0} = \frac{1}{\sqrt{2}}(\ket{0,N}_{\text{m}} +i \ket{z,N}_{\text{m}} )$.


We calculate the uncertainty of the estimation with the spin cat state under the effect of dephasing
when the applied field is aligned to $\vec n = (0,0,1)$.
If the memory qubits are prepared in a spin coherent state $\ket{z ,N}$,
the mixed state after the dephasing is described by 
\begin{eqnarray}
\tilde{\rho} = {\mathcal N}_z \left[ |g\rangle \langle g|+\tilde{z}^*|g\rangle \langle e|+\tilde{z}|e\rangle \langle g|+|z|^2|e\rangle \langle e| \right]^{\otimes N}
\end{eqnarray}
where $\tilde{z}=z e^{i\omega t-(\Gamma _tt)^2}$ with ${\mathcal N}_z = (1 + |z|^2 )^{-N}$.
For the spin cat state, we find
 \begin{eqnarray}
 \rho_t \propto |0\rangle\langle 0|+\tilde{\rho} + \mathcal{N}_d \left[ |\tilde{z}\rangle\langle 0|+|0\rangle\langle \tilde{z}| \right] , 
 \end{eqnarray}
where
$\mathcal{N}_d = (1+|z|^2e^{-2(\Gamma_tt)^2})^{N/2}/(1+|z|^2)^{N/2}$.
Then, by choosing the estimator to be
$A = \ket {\Phi _0} \bra {\Phi _0}$,
we obtain $ \delta \omega \simeq \frac{1+|z|^2}{Nt|z|^2}\sqrt{\frac{t}{T}}$
for $N\omega t\ll 1$ and $N(\Gamma _tt)^2\ll 1$.
In order to satisfy the condition $N(\Gamma_tt)^2\ll 1$,
we choose $t=\frac{s}{\gamma \sqrt{N}}$ (where $s$ denotes a small number )
and obtain
\begin{eqnarray}\label{eq_spincat}
 \delta \omega \simeq \frac{1+|z|^2}{|z|^2}\sqrt{\frac{\gamma }{sT}}N^{-\frac{3}{4}}.
\end{eqnarray}
This beats the SQL in scaling as long as $|z|=O(N^0)$.

\begin{figure}[tb]
\centering
\includegraphics[width=80mm, height=40mm]{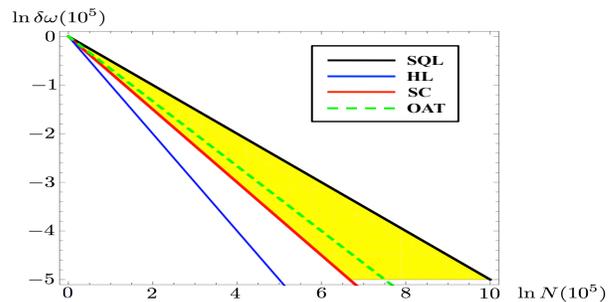}
\caption{(Color Online).  Scalings of the estimation uncertainty
$\delta \omega $ with respect to the particle number of memory qubits.
Black and blue lines denotes the scalings of the SQL and Heisenberg limit, respectively.
{The estimation uncertainty with spin cat states varies in the yellow
 area, depending on the value of $|z|$.}
Here, we set $|z| \sim N^k (k \leq 0)$ and $\sqrt{\gamma /(sT)} = 1$ in Eq.~(\ref{eq_spincat}).
The red line denotes the best scaling of $\delta \omega$ with spin cat
 states or the two-axis twisted spin squeezed states under the effect of non-Markovian dephasing.
The green line denotes the scaling of $\delta \omega$ with the one-axis twisted spin squeezed states under the effect of non-Markovian dephasing.
 }

\label{Relation}
\end{figure}

Phase measurement with a cat state has been discussed in optics~\cite{ralph2002coherent,joo2011quantum,knott2014attaining}.
However, the optical cat state is fragile against photon
loss~\cite{mark_noise_2,mark_noise_3,mark_noise_4}, which may provide a limitation for the practical application of the spin
cat state in optics.
On the other hand, in solid-state systems, the main decoherence is non-Markovian dephasing~\cite{noise_solid_1,noise_solid_2,noise_solid_3,noise_solid_4,noise_solid_5,noise_solid_6,noise_solid_7,noise_solid_8,noise_solid_9},
and this fact makes the magnetic field sensor with the spin cat state quite robust against experimental imperfections, as described above.
So, unlike the conventional expectations in the field of optics,
we have succeeded in showing that the spin cat state would provide us with quantum advantage to beat the SQL in scaling.


{{\it Entanglement sensor with spin squeezed states--}}
There are many protocols to generate a spin
squeezed state by global operations~\cite{sc_shane2014,squeezed_review}.
For example, one-axis twisted spin squeezed states are generated by using the flip-flop type
interaction defined by $H_I^{(2)}$~\cite{sc_shane2014}, which provides us with the ability to
squeeze the memory qubits via the collective interaction with the
control qubit.
Although it is experimentally more difficult to generate the two-axis twisted state,
one can in principle generate this state by successive application of one-axis twisting and microwave pulses to memory qubits~\cite{cre_TAT_1,cre_TAT_2,cre_TAT_3}.

A quantum state $\ket{\psi}$ is called spin squeezed~\cite{wineland1992,wineland1994} if the following inequality holds:
\begin{equation}
\xi^2_W := \min_{\vec r \ : \ \vec r \cdot \vec m = 0 } \frac{N \cdot {\mbox{Var}}_{\psi} (J_{ \vec r })}{[\mbox{E}_{\psi} (J_{\vec m})]^2} <1,
\end{equation}
where $\vec m$ is a mean spin vector
defined by
$(\mbox{E}_{\psi}(J_x) , \mbox{E}_{\psi}(J_y) , \mbox{E}_{\psi}(J_z) )$.
{This definition is derived from
a sensitivity to measure the phase without decoherence when a collective angular momentum
operator is
an estimator.}
Note that for the spin coherent state $\xi_W^2 =1$, while for the one-axis twisted state $\xi_W^2 = O(N^{-\frac{2}{3}})<1$.

We consider the field sensitivity of spin squeezed states under the effect of non-Markovian dephasing.
In our setup, {to obtain the phase information acquired in the
memory qubits during the exposure to the magnetic field,}
we read out an expectation value of a specific collective angular momentum
operator $J_{\vec r}^{(\text{m})}$ via {the application
of the resonant microwave pulse on the memory qubits, the interaction Hamiltonian $H_I^{(1)}$, and a projective measurement on the control qubit.
To use the spin squeezed state for the sensing, we need to choose
(i) an angular momentum operator for the estimator
and
(ii) the sensing direction.}
{For (i), in order to avoid quantum fluctuations during the measurement process,}
we choose the vector $\vec r$ to give the minimal variance of a collective angular operator for the entangled state $\ket \psi$.
For (ii),
we choose the sensing direction to maximize the variance of $J_{\vec n}^{(\text{m})}$ for the entangled state $\ket \psi$.
This is because the quantum Fisher information is proportional to the variance of $J_{\vec n}^{(\text{m})}$ without the effect of decoherence.
From these, we can calculate the estimation uncertainty:
\begin{equation}\label{eq:t01}
\delta \omega =\sqrt{ \frac{{\mbox{Var}}_{\psi} (J_{ \vec r })  + N(e^{2 (\Gamma_t t)^2} -1)/4+ \Delta _1 + \Delta_2}{Tt [{\mbox{E}}_{\psi} (J_{ \vec m }) ]^2}},
\end{equation}
with $\Delta_1 \leq 0$ and $\Delta_2 = a (\tan(\omega t) + b)^2$.
Here, $a$ ($b$) is positive (real) and independent of $\omega$~\cite{footnote_3}.

Next, we evaluate an upper bound for $\delta \omega$. 
First, from equation (\ref{eq:t01}), we have $\delta \omega \leq f(t)$ with $\omega t$ satisfying ${\mbox{Var}}_{\psi} (J_{ \vec r })  + N(e^{2 (\Gamma_t t)^2} -1)/4 \geq \Delta_2$, where
\begin{equation}\label{eq:t04}
f(t) =  \sqrt{\frac{2\{ {\mbox{Var}}_{\psi} (J_{ \vec r })  + N(e^{2 (\Gamma_t t)^2} -1)/4\}}{Tt [{\mbox{E}}_{\psi} (J_{ \vec m }) ]^2}} .
\end{equation}
Then, by setting the exposure time $t$ as $t = \alpha N^{-s_1}$ with $\alpha \geq 0$ and $s_1 \geq 0$,
in the large $N$ limit (short time limit) the upper bound of $\delta \omega$
is simplified to
\begin{equation}
f(\alpha N^{-s_1}) = \sqrt{\frac{N^{s_1}}{\alpha T}\frac{2 {\mbox{Var}}_{\psi} (J_{ \vec r } ) + \beta \gamma ^2  N^{-2 s_1+1} }{ [{\mbox{E}}_{\psi} (J_{ \vec m }) ]^2}} ,
\end{equation}}
with $\beta$ positive and independent of $N$.

Finally, we determine the scaling of $f(N)$.
Notice that, since ${\mbox{E}}_{\psi} (J_{ \vec m })$ (${\mbox{Var}}_{\psi} (J_{ \vec r })$) is the first (second) moment of the collective angular operator, we can set its order as
$
 {\mbox{E}}_{\psi} (J_{ \vec m }) = O(N^{s_2}) \ ( {\mbox{Var}}_{\psi} (J_{ \vec r } )  =O( N^{s_3}) )
$
with $0 \leq s_2 \leq 1$ ($0 \leq s_3 \leq 2$ ).
Then, we have
\begin{equation}
f(N) = \begin{cases}O(N^{\frac{s_1 + s_3}{2} -s_2}) \quad \mbox{for} \ s_3 > 1, \\
 O(N^{\frac{1-s_1}{2} - s_2}) \quad \mbox{for} \ s_1 < (1-s_3)/2,  \\
O(N^{\frac{s_1 + s_3}{2} -s_2}) \quad \mbox{for} \  s_1 \geq (1-s_3)/2 .
\end{cases} 
\end{equation}
The smallest scaling of $f(N)$ is obtained when $s_3 \leq 1$ and $s_1 =
(1-s_3)/2$, and
the uncertainty of estimation is bounded by
$
\delta \omega \leq O(N^{-s_2 + \frac{1+s_3}{4}}) .
$
Especially, the expectation value of the mean-spin-directed collective angular operator is usually proportional to the particle number, i.e., $s_2 =1$.
Then, we have $\delta  \omega \leq O(N^{\frac{-3 +s_3}{4} })$.
This means that
the smaller the variance of an estimator is, the more precisely we can estimate the magnetic field.
Moreover, whenever the variance of an estimator is smaller than the particle number, i.e., $s_3 < 1$,
we can beat the SQL in scaling even under non-Markovian dephasing.

Let us give examples of spin squeezed states that beat the SQL in scaling.
When the memory qubits are prepared in the one-axis twisted state, we
have ${\mbox{E}}_{\psi} (J_{ \vec m }) = O(N^1)$ and
${\mbox{Var}}_{\psi} (J_{ \vec r }) =
O(N^{\frac{1}{3}})$~\cite{squeezed_spin:1993}, and the uncertainty is bounded by $\delta \omega \leq O(N^{-\frac{2}{3}})$.
Another example is a two-axis twisted state.
For this state,
we have ${\mbox{E}}_{\psi} (J_{ \vec m }) = O(N^1)$ and ${\mbox{Var}}_{\psi} (J_{ \vec r }) = O(N^0)$~\cite{squeezed_spin:1993},
and thus an uncertainty
$\delta \omega \leq O(N^{-\frac{3}{4}})$ is achieved.

Although there are established techniques to generate spin squeezed states,
the accepted belief has been that one cannot beat the SQL with the spin
squeezed state under decoherence \cite{squeezed_review}.
In particular, previous authors have estimated the effect of Markovian dephasing and shown that
the sensitivity of the spin squeezed sensor is the same as the SQL~\cite{kitagawa2001}.
However, we have found that,
under the effect of non-Markovian dephasing---the dominant source of decoherence in solid state systems---the spin squeezed state can measure a magnetic field with a sensitivity which beats the SQL in scaling.

{{\it Summary--}}
Spin cat states and spin squeezed states are considered to be experimentally feasible entangled states,
because one can create these via global operations without individual
control of qubits. We firstly analyze the sensitivity of a quantum
sensor using such entanglement
in the presence of general non-Markovian
phase noise.
We show that, by using these entangled states, one can sense the magnetic field with an accuracy far beyond the classical sensor
even under the effect of realistic decoherence. 
Our results pave the way for the practical implementation of magnetic field sensors that can exploit entanglement to operate below the limits of classical physics.
T.T thanks H. Nakazato and K. Yuasa for discussion.
This work was supported in part
by
Commissioned Research of NICT.


\begin{thebibliography}{99}

\bibitem{metro_bio}
M. A. Taylor and W. P. Bowen, arXiv:1409.0950.

\bibitem{review_q_metro_1}
V. Giovannetti, S. Lloyd and L. Maccone, Science {\bf 306}, 1330 (2004).

\bibitem{review_q_metro_2}
V. Giovannetti, S. Lloyd and L. Maccone, Nature Photon {\bf 5}, 222 (2011).

\bibitem{wineland1992}
D. J. Wineland, J. J. Bollinger, W. M. Itano, F. L. Moore and D. J. Heinzen, Phys. Rev. A {\bf 46}, R6797 (1992).

\bibitem{wineland1994}
D. J. Wineland, J. J. Bollinger, W. M. Itano and D.J. Heinzen, Phys. Rev. A {\bf 50}, 67 (1994).

\bibitem{review_q_metro_3}
M. G. A. Paris, Int. J. Quant. Inf. {\bf 07}, 125 (2009).

\bibitem{review_q_metro_4}
C. Gross, J. Phys. B: At. Mol. Opt. Phys. {\bf 45}, 103001 (2012).

\bibitem{review_q_metro_5}
G. T\'oth and I. Apellaniz, J. Phys. A: Math. Theor. {\bf 47} 424006 (2014).

\bibitem{mark_noise_1}
S. F. Huelga, C. Macchiavello, T. Pellizzari, A. K. Ekert, M. B. Plenio and J. I. Cirac, Phys. Rev. Lett. {\bf 79}, 3865 (1997).

\bibitem{mark_noise_2}
B. M. Escher, R. L. de Mantos, and L. Davidovich, Nature Phys., {\bf 7}, 406 (2011).

\bibitem{mark_noise_3}
 R. Demkowicz-Dobrza{\'n}ski, J. Ko{\l}ody{\'n}ski, and M. Gu{\c{t}}{\u{a}}, Nature Commun., {\bf 3}, 1063 (2012).

\bibitem{mark_noise_4}
J. Ko{\l}ody{\'n}ski and R. Demkowicz-Dobrza{\'n}ski, New J. Phys., {\bf 15}, 073043 (2013).

\bibitem{mark_noise_5}
R. Chaves, J. B. Brask, M. Markiewicz, J. Ko{\l}ody{\'n}ski, and A. Ac{\'i}n, Phys. Rev. Lett. {\bf 111}, 120401 (2013).

\bibitem{non_mark_noise_1}
Y. Matsuzaki, S. C. Benjamin and J. Fitzsimons, Phys. Rev. A {\bf 84}, 012103 (2011).

\bibitem{non_mark_noise_2}
A. W. Chin, S. F. Huelga and M. B. Plenio, Phys. Rev. Lett. {\bf 109}, 233601 (2012).

\bibitem{noise_solid_1}
E. Paladino, Y. M. Galperin, G. Falci and B. L. Altshuler, Rev. Mod. Phys. {\bf 86}, 361 (2014).

\bibitem{noise_solid_2}
G. de Lange, Z. H. Wang, D. Rist\`e, V. V. Dobrovitski and R. Hanson, Science {\bf 330}, 60 (2010).

\bibitem{noise_solid_3}
P. L. Stanwix, L. M. Pham, J. R. Maze, D. Le Sage, T. K. Yeung, P. Cappellaro, P. R. Hemmer, A. Yacoby, M. D. Lukin and R. L. Walsworth, Phys. Rev. B {\bf 82}, 201201(R) (2010).

\bibitem{noise_solid_4}
J. R. Maze, J. M. Taylor and M. D. Lukin, Phys. Rev. B {\bf 78}, 094303 (2008).

\bibitem{noise_solid_5}
K. Kakuyanagi, T. Meno, S. Saito, H. Nakano, K. Semba, H. Takayanagi, F. Deppe and A. Shnirman, Phys. Rev. Lett. {\bf 98}, 047004 (2007).

\bibitem{noise_solid_6}
F. Yoshihara, K. Harrabi, A. O. Niskanen, Y. Nakamura and J. S. Tsai, Phys. Rev. Lett. {\bf 97}, 167001 (2006).

\bibitem{noise_solid_7}
J. Bylander, S. Gustavsson, F. Yan, F. Yoshihara, K. Harrabi, G. Fitch, D. G. Cory, Y. Nakamura, J. Tsai and W. D. Oliver, Nature Phys., {\bf 7}, 565 (2011).

\bibitem{noise_solid_8}
E. Paladino, L. Faoro, G. Falci, and R. Fazio, Phys. Rev. Lett. {\bf 88}, 228304 (2002).

\bibitem{noise_solid_9}
B. E. Kane, Nature {\bf 393}, 133 (1998).

\bibitem{cre_ghz_1}
D. Leibfried, M. D. Barrett, T. Schaetz, J. Britton, J. Chiaverini, W. M. Itano, J. D. Jost, C. Langer and D. J. Wineland, Science {\bf 304}, 1476 (2004).

\bibitem{cre_ghz_2}
T. Monz, P. Schindler, J. T. Barreiro, M. Chwalla, D. Nigg, W. A. Coish, M. Harlander, W. H\"ansel, M. Hennrich and R. Blatt, Phys. Rev. Lett. {\bf 106}, 130506 (2011).

\bibitem{cre_ghz_3}
R. Barends, J. Kelly, A. Megrant, A. Veitia, D. Sank, E. Jeffrey, T. C. White, J. Mutus, A. G. Fowler, B. Campbell, Y. Chen, Z. Chen, B. Chiaro, A. Dunsworth, C. Neill, P. O'Malley, P. Roushan, A. Vainsencher, J. Wenner, A. N. Korotkov, A. N. Cleland and J. M. Martinis, Nature {\bf 508}, 500 (2014).


\bibitem{sc_shane2014}
S. Dooley and T. P. Spiller, Phys. Rev. A {\bf 90}, 012320 (2014).

\bibitem{sc_shane2014_2}
S. Dooley, J. Joo, T. Proctor and T. P. Spiller, arXiv:1406.6036.

\bibitem{squeezed_review}
J. Ma, X. G. Wang, C. P. Sun, and F. Nori, Phys. Rep. {\bf 509}, 89 (2011).

\bibitem{squeezed_experi_1}
K. Hammerer, A. S. S{\o}rensen and E. S. Polzik, Rev. Mod. Phys. {\bf 82}, 1041 (2010).

\bibitem{squeezed_spin:1993}
M. Kitagawa and M. Ueda, Phys. Rev. A {\bf 47}, 5138 (1993).

\bibitem{footnote_1}
\textcolor{black}{Take $\Gamma_t = 0$ in Eq.~(\ref{noise_map}).
Then, our results produce a well known fact that the scaling of the estimation uncertainty without noise effects is $O(N^{-1})$ for spin cat and two-axis twisted states,
while it is $O(N^{-\frac{5}{6}})$ for the one-axis twisted state.
\textcolor{black}{For details of the estimation uncertainty of spin cat states without noise effects, see Refs.~\cite{qft_spin_cat,shanecat2013}.}}

\bibitem{qft_spin_cat}
H. Xiong, J. Ma, W. Liu and X. Wang, Quantum Inf. Comput. {\bf 10}, 498 (2010).

\bibitem{shanecat2013}
S. Dooley, F. McCrossan, D. Harland, M. J. Everitt and T. P. Spiller, Phys. Rev. A {\bf 87}, 052323 (2013).


\bibitem{marcos2010coupling}
D. Marcos, M. Wubs, J. M. Taylor, R. Aguado, M. D. Lukin and A. S. S{\o}rensen, Phys. Rev. Lett. {\bf 105}, 210501 (2010).

\bibitem{zhu2011coherent}
X. Zhu, S. Saito, A. Kemp, K. Kakuyanagi, S. Karimoto, H. Nakano, W. Munro, Y. Tokura, M. Everitt, K. Nemoto, M. Kasu, N. Mizuochi and K. Semba, Nature {\bf 478}, 221 (2011).

\bibitem{saito2013towards}
S. Saito, X. Zhu, R. Ams\"uss, Y. Matsuzaki,
K. Kakuyanagi, T. Shimo-Oka, N. Mizuochi, K. Nemoto, W. J. Munro and K. Semba, Phys. Rev. Lett. {\bf 111}, 107008 (2013).

\bibitem{ClarkeWilhelm01a}
J. Clarke and F. Wilhelm, Nature {\bf 453}, 1031 (2007) 

\bibitem{footnote_M}
An NV center
is not a two-level system but a three-level system because of the spin
1 structure. However, 
applying magnetic field induces the Zeeman splitting to isolate a
two-level subsystem of the NV center so that we can treat the NV center
as an effective qubit.

\bibitem{maurer2012room}
P. Maurer, G. Kucsko, C. Latta, L. Jiang, N. Yao, S. Bennett, F. Pastawski, D. Hunger, N. Chisholm, M. Markham, D. J. Twitchen, J. I. Cirac and M. D. Lukin, Science {\bf 336}, 1283 (2012).

\bibitem{footnote_2}
In this paper, the expectation value of an observable $A$ under a state $\rho$ is written as
${\mbox{E}}_{\rho}(A) = {\mbox{tr}}(A \rho)$, while the variance is written as
${\mbox{Var}}_{\rho}(A) = {\mbox{tr}}(A^2 \rho) -{\mbox{E}}_{\rho}(A) ^2 $.
In addition, if a state is a pure state ($\rho = \ket \psi \bra \psi$), we use the following shorthand notations. ${\mbox{E}}_{\psi}(A) = {\mbox{tr}}(A \ket \psi \bra \psi)$ and ${\mbox{Var}}_{\psi}(A) = {\mbox{tr}}(A^2 \ket \psi \bra \psi) -{\mbox{E}}_{\psi}(A) ^2 $.



\bibitem{PhysRevA.70.022327}
M. Hotta and M. Ozawa, Phys. Rev. A {\bf 70}, 022327 (2004).

\bibitem{qft_ref_2}
W. Zhong, X. M. Lu, X. X. Jing and X. Wang, J. Phys. A: Math. Theor. {\bf 47} 385304 (2014).


\bibitem{ralph2002coherent}
T. C. Ralph, Phys. Rev. A {\bf 65}, 042313 (2002).

\bibitem{joo2011quantum}
J. Joo, W. J. Munro and T. P. Spiller, Phys. Rev. Lett. {\bf 107}, 083601 (2011).

\bibitem{knott2014attaining}
P. A. Knott, T. J. Proctor, K. Nemoto, J. A. Dunningham and W. J. Munro, Phys. Rev. A {\bf 90}, 033846 (2014).


\bibitem{cre_TAT_1}
Y. C. Liu, Z. F. Xu, G. R. Jin and L. You, Phys. Rev. Lett. {\bf 107}, 013601 (2011).

\bibitem{cre_TAT_2}
C. Shen and L. M. Duan, Phys. Rev. A {\bf 87}, 051801(R) (2013).

\bibitem{cre_TAT_3}
J. Zhang, X. Zhou, G. Guo and Z. Zhou, Rev. A {\bf 90}, 013604 (2014).


\bibitem{footnote_3}
$\Delta_1$, $a$ and $b$ are given as follows.
$\Delta_1 = - \{ {\mbox{E}}_{\psi}(J_{\vec r} J_{\vec m} + J_{\vec m} J_{\vec r} )\}^2/\{ 4{\mbox{Var}}_{\psi}(J_{\vec m}) + N(e^{2 (\Gamma_t t)^2} -1) \}$, $a = {\mbox{Var}}_{\psi}(J_{\vec m}) + N(e^{2 (\Gamma_t t)^2} -1)/4$, and $b=\{ {\mbox{E}}_{\psi}(J_{\vec r} J_{\vec m} + J_{\vec m} J_{\vec r} )\} /\{ 2{\mbox{Var}}_{\psi}(J_{\vec m}) + N(e^{2 (\Gamma_t t)^2} -1)/2 \}$.

\bibitem{kitagawa2001}
D. Ulam-Orgikh and M. Kitagawa, Phys. Rev. A {\bf 64}, 052106 (2001).






\end{thebibliography}
\end{document}